\documentclass[a4paper,fleqn,usenatbib,useAMS]{mnras}

\usepackage[T1]{fontenc}
\usepackage{ae,aecompl}
\usepackage{float}
\usepackage{xcolor}

\usepackage{graphics,graphicx}	
\usepackage{amsmath}	
\usepackage{amssymb}	

\usepackage{newtxtext,newtxmath}

\usepackage{color,hyperref}
\hypersetup{colorlinks,breaklinks,
            linkcolor=blue,urlcolor=blue,
            anchorcolor=blue,citecolor=blue}

\usepackage{caption}
\usepackage{longtable}
\DeclareCaptionFormat{cont}{#1 (continued.)#2#3\par}

\title{The \textit{Gaia} alerted fading of the FUor-type star Gaia21elv}

\author[Z. Nagy et al.]{
Zs\'ofia Nagy,$^{1,2}$\thanks{E-mail: nagy.zsofia@csfk.org}
Sunkyung Park,$^{1,2}$
P\'{e}ter \'{A}brah\'{a}m,$^{1,2,3}$
\'{A}gnes K\'{o}sp\'{a}l,$^{1,2,3,4}$ 
Fernando Cruz-S\'aenz de Miera,$^{1,2}$
\newauthor
M\'aria Kun,$^{1,2}$
Micha{\l} Siwak,$^{1,2}$
Zs\'ofia Marianna Szab\'o,$^{1,2,5,6}$
M\'{a}t\'{e} Szil\'{a}gyi,$^{1,2,3}$
Eleonora Fiorellino,$^{7}$
\newauthor
Teresa Giannini,$^{8}$
Jae-Joon Lee,$^{9}$
Jeong-Eun Lee,$^{10}$
G\'abor Marton,$^{1,2}$
L\'aszl\'o Szabados,$^{1,2}$
Fabrizio Vitali,$^{8}$
\newauthor
Jan Andrzejewski,$^{11}$
Mariusz Gromadzki,$^{12}$
Simon Hodgkin,$^{13}$
Maja Jab{\l}o{\'n}ska,$^{12}$
Rene A. Mendez,$^{14}$
\newauthor
Jaroslav Merc,$^{15}$
Olga Michniewicz,$^{11}$
Przemys{\l}aw J. Miko{\l}ajczyk,$^{12,16}$
Uliana Pylypenko,$^{12}$
\newauthor
Milena Ratajczak,$^{12}$
{\L}ukasz Wyrzykowski,$^{12}$
Michal Zejmo,$^{11}$
Pawe{\l} Zieli{\'n}ski$^{17}$ \\
$^{1}$Konkoly Observatory, Research Centre for Astronomy and Earth Sciences, E\"otv\"os Lor\'and Research Network (ELKH), \\ H-1121 Budapest, Konkoly Thege Mikl\'os \'ut 15-17., Hungary \\
$^{2}$CSFK, MTA Centre of Excellence, Konkoly-Thege Mikl\'os \'ut 15-17, 1121 Budapest, Hungary \\
$^{3}$ELTE E\"otv\"os Lor\'and University, Institute of Physics, P\'azm\'any P\'eter s\'et\'any 1/A, H-1117 Budapest, Hungary\\
$^{4}$Max Planck Institute for Astronomy, K\"onigstuhl 17, D-69117 Heidelberg, Germany \\
$^{5}$Max Planck Institute for Radio Astronomy, Auf dem Hügel 69, 53121 Bonn, Germany \\
$^{6}$Scottish Universities Physics Alliance (SUPA), School of Physics and Astronomy, University of St Andrews, North Haugh, St Andrews, KY16 9SS, UK \\
$^{7}$INAF-Osservatorio Astronomico di Capodimonte, via Moiariello 16, 80131 Napoli, Italy\\
$^{8}$INAF-Osservatorio Astronomico di Roma, via di Frascati 33, 00078, Monte Porzio Catone, Italy\\
$^{9}$Korea Astronomy and Space Science Institute 776, Daedeok-daero, Yuseong-gu, Daejeon, 34055, Republic of Korea \\
$^{10}$Department of Physics and Astronomy, Seoul National University, 1 Gwanak-ro, Gwanak-gu, Seoul 08826, Republic of Korea \\
$^{11}$Janusz Gil Institute of Astronomy University of Zielona Gora, Poland \\
$^{12}$Warsaw University Observatory, Al. Ujazdowskie 4, 00-478 Warsaw, Poland\\
$^{13}$Institute of Astronomy, Madingley Road, Cambridge CB3 0HA, UK \\
$^{14}$Astronomy Department, Universidad de Chile, Casilla 36-D, Santiago, Chile\\
$^{15}$Astronomical Institute, Faculty of Mathematics and Physics, Charles University, V Hole\v{s}ovi\v{c}k{\'a}ch 2, 180 00 Prague, Czech Republic\\
$^{16}$Astronomical Institute, University of Wroc{\l}aw, ul. M. Kopernika 11, 51-622 Wroc{\l}aw, Poland \\
$^{17}$Institute of Astronomy, Faculty of Physics, Astronomy and Informatics, Nicolaus Copernicus University in Toru{\'n},\\ ul. Grudzi\k{a}dzka 5, 87-100 Toru{\'n}, Poland \\
}

\date{Accepted XXX. Received YYY; in original form ZZZ}

\pubyear{2023}
 
\begin{document}
\label{firstpage}
\pagerange{\pageref{firstpage}--\pageref{lastpage}}
\maketitle
 
\begin{abstract}
FU Orionis objects (FUors) are eruptive young stars, which exhibit outbursts that last from decades to a century. Due to the duration of their outbursts, and to the fact that only about two dozens of such sources are known, information on the end of their outbursts is limited. 
Here we analyse follow-up photometry and spectroscopy of Gaia21elv, a young stellar object, which had a several decades long outburst. It was reported as a \textit{Gaia} science alert due to its recent fading by more than a magnitude. To study the fading of the source and look for signatures characteristic of FUors, we have obtained follow-up near infrared (NIR) spectra using Gemini South/IGRINS, and both optical and NIR spectra using VLT/X-SHOOTER. The spectra at both epochs show typical FUor signatures, such as a triangular shaped $H$-band continuum, absorption-line dominated spectrum, and P Cygni profiles. In addition to the typical FUor signatures, [\ion{O}{i}], [\ion{Fe}{ii}], and [\ion{S}{ii}]~were detected, suggesting the presence of a jet or disk wind. 
Fitting the spectral energy distributions with an accretion disc model suggests a decrease of the accretion rate between the brightest and faintest states. The rapid fading of the source in 2021 was most likely dominated by an increase of circumstellar extinction.
The spectroscopy presented here confirms that Gaia21elv is a classical FUor, the third such object discovered among the \textit{Gaia} science alerts.
\end{abstract}



\section{Introduction}

Studying the accretion in young stellar objects (YSOs) is important to understand their formation. 
Most of what we know about accretion in YSOs is based on the magnetospheric accretion scenario, according to which the material accretes onto the forming star from the infalling envelope through the disk, by following the magnetospheric lines \citep{Hartmann2016}.
The accretion rates of YSOs are known to be highly variable, with extreme cases of eruptive YSOs, which experience outburst events, when their luminosity increases up to two orders of magnitude. These events are detected as 2-5 mag brightening in optical and near-infrared (NIR) bands. 
During the outbursts the mass accretion rate can increase from $\sim$10$^{-8}$ $M_\odot$ yr$^{-1}$ in quiescence to $\sim$10$^{-4}$ $M_\odot$ yr$^{-1}$ (\citealp{Audard2014}, \citealp{Fischer2022}). Studies with large samples of objects indicate that young stars experience these events once every $10^3-10^4$ years (e.g. \citealp{Fischer2019}). 
Episodic accretion is one of the possible explanations for the observed large luminosity spread of young stellar objects \citep{Fischer2022}. 
FU Orionis objects (FUors) are well-studied examples of episodic accretion \citep{HartmannKenyon1996}. FUors are low-mass ($<2~M_\odot$) eruptive YSOs that exhibit large-amplitude ($>$4 mag) outbursts at optical and infrared wavelengths. These outbursts are expected to last up to a century, suggesting that these events will not only increase the final stellar mass by a significant amount, but also affect the evolution of the circumstellar disc. The representative characteristics of FUors are brightness increase on a time scale of 1-10 yr, P Cygni profile of H$\alpha$, \ion{Li}{i} 6707 \AA~absorption, strong CO absorption features, triangular shape of the $H$-band continuum due to the strong water absorption bands on both sides of the $H$-band window,
typical of late M-type stars (\citealp{HartmannKenyon1996}; \citealp{ConnelleyReipurth2018}). So far the number of confirmed FUors is limited to no more than two dozens \citep{Audard2014}.
One of the important, so far unclear points is the end of the FUor outbursts, i.e. their return to quiescence. FUor outbursts are expected to end when the inner disc depletes.
However, due to the typically decades-long duration of the outbursts, no bona fide FUor has returned to quiescence yet, apart from cases of short, temporary halt in the accretion, e.g. V899 Mon \citep{Ninan2015,Park2021} and V346 Nor (\citealp{Kraus2016}, \citealp{Kospal2020}). 
Another example is V1647 Ori, an eruptive YSO that has shown some FUor characteristics \citep{Aspin2009}, and returned to quiescence after a ten-years long outburst \citep{Semkov2018,Giannini2018}. The spectroscopic deviation of V1647 Ori from well-known FUors, however, ruled out its FUor classification \citep{ConnelleyReipurth2018}.  

Therefore, it is not known whether the end of FUor outbursts is an abrupt event when accretion suddenly stops and the brightness drops back to the quiescent level in 1-2 years, or it is a slow gradual decrease of the accretion rate resulting in a slowly decreasing light curve over perhaps decades. The first scenario would indicate some instability, like the thermal instability model proposed by \citet{Bell1994}. To understand how FUors end their outbursts, it is important to increase their sample.

One of the best tools to discover the brightening or fading of eruptive young star candidates is the \textit{Gaia} Photometric Science Alerts system, due to its large sky coverage and typically monthly cadence \citep{Hodgkin2021}. Several eruptive YSOs have already been discovered based on the \textit{Gaia} Science Alerts, including the FUors Gaia17bpi \citep{Hillenbrand2018} and Gaia18dvy \citep{SzegediElek2020}, and the EX Lupi-type eruptive YSOs (EXors) Gaia18dvz \citep{Hodapp2019}, Gaia20eae \citep{Ghosh2022, CruzSaenzdeMiera2022} and Gaia19fct \citep{Park2022}. 
Some additional eruptive YSOs were found, which cannot be classified as either a FUor or an EXor, such as Gaia19ajj \citep{Hillenbrand2019}, Gaia19bey \citep{Hodapp2020}, and Gaia21bty (Siwak et al., submitted).
Two \textit{Gaia} alerted sources with light curves similar to eruptive YSOs, Gaia20bwa and Gaia20fgx \citep{Nagy2022}, turned out to be classical T Tauri stars (CTTS), while the brightening of another \textit{Gaia} alerted YSO, V555 Ori (Gaia17afn), was confirmed to be caused by variable circumstellar extinction, rather than a change in its accretion rate \citep{Nagy2021}. Here we present a study of a previously known YSO, which triggered the \textit{Gaia} Science Alerts system due to its fading.

Gaia21elv (ESO H$\alpha$-148 or 2MASS J08410676-4052174, $\alpha_{\rm J2000}$ = 08$^{\rm h}$ 41$^{\rm m}$ 06$\fs$75, $\delta_{\rm J2000}$ = -40$^{\circ}$ 52$'$ 17$\farcs$44) had a \textit{Gaia} alert on 2021 October 6 due to its quick fading by 1.2 mag over 18 months. Its archival photometry based on photographic plates of the SuperCOSMOS Sky Survey (SSS) showed a long-term brightening \citep{ContrerasPena2019}.
It is a known young, Class II type star (\citealp{PetterssonReipurth1994}, \citealp{Marton2019}), associated with the Vela Molecular Ridge \citep{PetterssonReipurth1994}, and in particular, with the RCW 27 HII region located at a distance of $\sim$1 kpc \citep{PetterssonReipurth2008}.
Its \textit{Gaia} DR3 \citep{Vallenari2022} parallax is 1.0727$\pm$0.0397 mas. The Renormalised Unit Weight Error (RUWE) of 1.291 and the astrometric excess noise of 0.437 mas suggest that the astrometry is accurate. We derived a zero-point correction of $-$0.02513 based on \citet{Lindegren2021} for this parallax.  
After the zero-point correction, the \textit{Gaia} DR3 parallax can be converted to a distance of 910.9$\pm$33.7 pc, which we use in this paper. This distance is close to the estimate of $905^{+36}_{-26}$ pc by \citet{BailerJones2021}.

In this paper, we provide spectroscopic evidence that Gaia21elv is a FUor, and discuss the cause of its fading that triggered the \textit{Gaia} Alerts system. 
We describe the photometric and spectroscopic observations in Sect. \ref{sect_observations} and present their results in Sect. \ref{sect_results}. We analyse the FUor signatures in the NIR spectra in Sect. \ref{sect_discussion}, discuss the nature of the fading of the source, and provide a comparison to other similar sources. We summarize our main findings in Sect. \ref{sect_summary}.

\section{Observations}   
\label{sect_observations}

\subsection{Optical photometry}

In 2022 June, we obtained optical photometric observations of Gaia21elv with the 60-cm Ritchey-Chr\'etien Rapid Eye Mount (REM) telescope operated by the Italian National Institute for Astrophysics (INAF) at La Silla (Chile) using its ROS2 instrument, an optical imager operating at four simultaneous passbands (Sloan $g'r'i'z'$) with a field of view (FoV) of 9$\farcm$1$\times$9$\farcm$1 and pixel scale of 0$\farcs$58. Three images were taken per filter on four nights, 2022 June 5, 6, 8, and 9. After the usual bias and flat field correction, and removal of hot pixels, we obtained aperture photometry for Gaia21elv and about 15 comparison stars in the FoV. We selected the comparison stars from the APASS9 catalog \citep{henden2015} making sure that they are sufficiently constant in brightness (${\sigma}_V<0.08\,$mag). We calculated the $z$-band brightness of the comparison stars by plotting their spectral energy distribution (SED) using APASS9 $Bg'Vr'i'$ and 2MASS $JHK_{\rm{s}}$ magnitudes \citep{cutri2003} and interpolating between these points for the effective wavelength of the $z'$ filter, 1.05$\,\mu$m.
We used an aperture radius of 6 pixels (3$\farcs$5) and sky annulus between 20 and 40 pixels (11$\farcs$68 and 23$\farcs$36). Because all comparison stars were much bluer than Gaia21elv, in order to avoid introducing large uncertainties by extrapolation, we converted the instrumental magnitudes by averaging the calibration factors of all comparison stars without fitting a colour term. The results can be seen in Table~\ref{tab:phot1}.

Further observations of the target have been performed with REM between 2022 Oct 26 and 2023 Jan 4, during 12 nights. These observations, taken in Sloan $g'r'i'$ passbands, were uploaded to the BHTOM service.\footnote{BHTOM - Black Hole TOM: \texttt{www.bhtom.space}} 
40, 38 and 44 images were reduced in Sloan $g'r'i'$, respectively.

Photometric observations were obtained with the PROMPT6 telescope located at Cerro Tololo Inter-American Observatory in Chile. This telescope is a part of SkyNET robotic network and is supplied with FLI CCD camera with 15.1 $\times$ 15.1 arcmin field-of-view (2048 $\times$ 2048 pixels, 0.44 arcsec/pix). All 42 observations (14 frames per band) were taken in Johnson-Cousins $V$, $R$ and $I$ bands and uploaded to the BHTOM service, where they were reduced and converted to standard magnitudes (in APASS/V, APASS/r and APASS/i respectively). 

We obtained photometric observations with the 1.54m Danish telescope, located at La Silla, Chile.
The telescope is equipped with the CCD camera (E2V231-42) in the Cassegrain focus, cooled by liquid nitrogen. The FoV is 13.7 $\times$ 13.7 arcmin (2048 $\times$ 2048 pixels; pixel scale of 0.4 arcsec/pixel). 
The filters used were Johnson-Cousins $B V R_{\rm{c}} I_{\rm{c}}$. In all cases, the exposure time was 90 seconds.

We collected data using the 50cm CDK telescope equipped with a QHY268M pro camera. This telescope (ROTUZ) is part of the DeepSkyChile\footnote{\texttt{https://www.deepskychile.com/en}}, and belongs to the Janusz Gil Institute of Astronomy, University of Zielona Gora, Poland. We reduced the data by applying bias, dark, and flat correction using AstroImageJ software \citep{Collins2017}. The photometry was done using the BHTOM server.
The photometry done using the BHTOM server is based on the method described in \citet{Zielinski2019} and \citet{Zielinski2020}.

\begin{figure*}
\centering
\includegraphics[width=17cm, trim={0cm 1cm 0 0cm},clip]{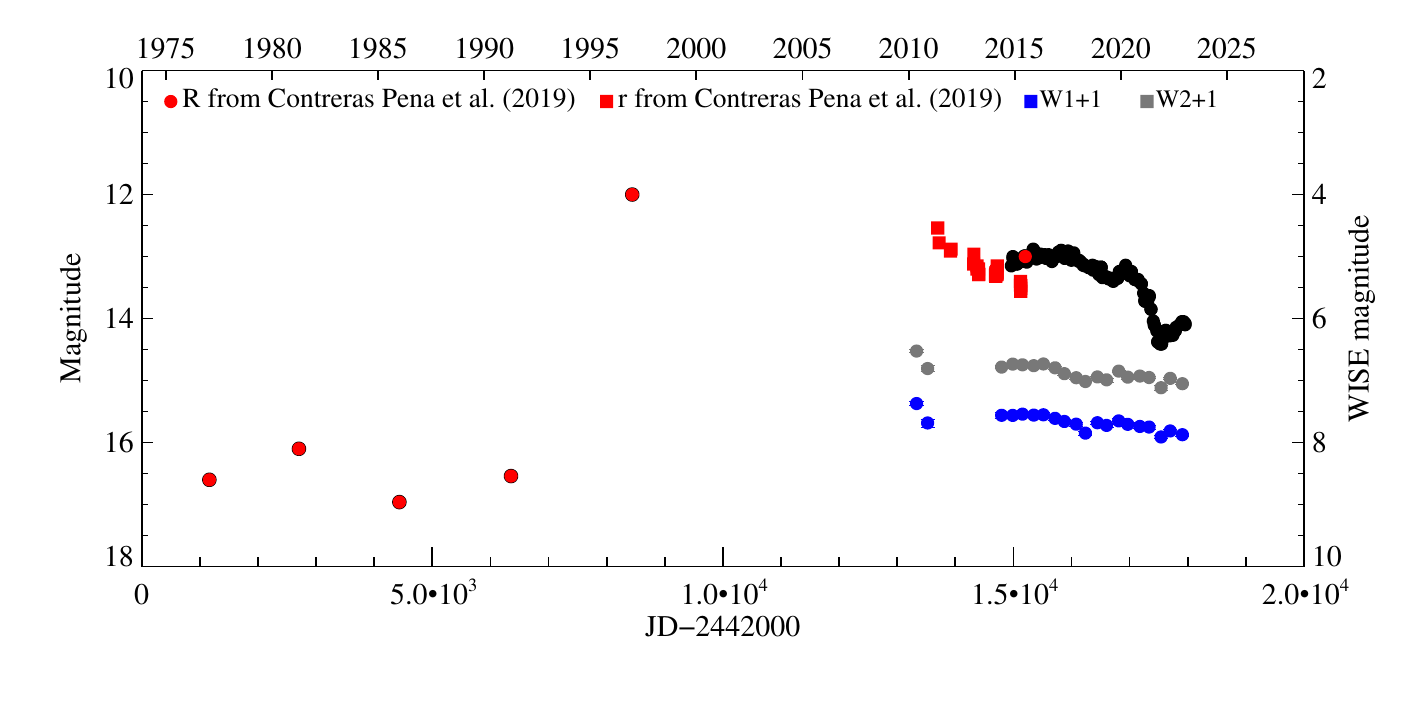}
\includegraphics[width=17cm, trim={0cm 1cm 0 0cm},clip]{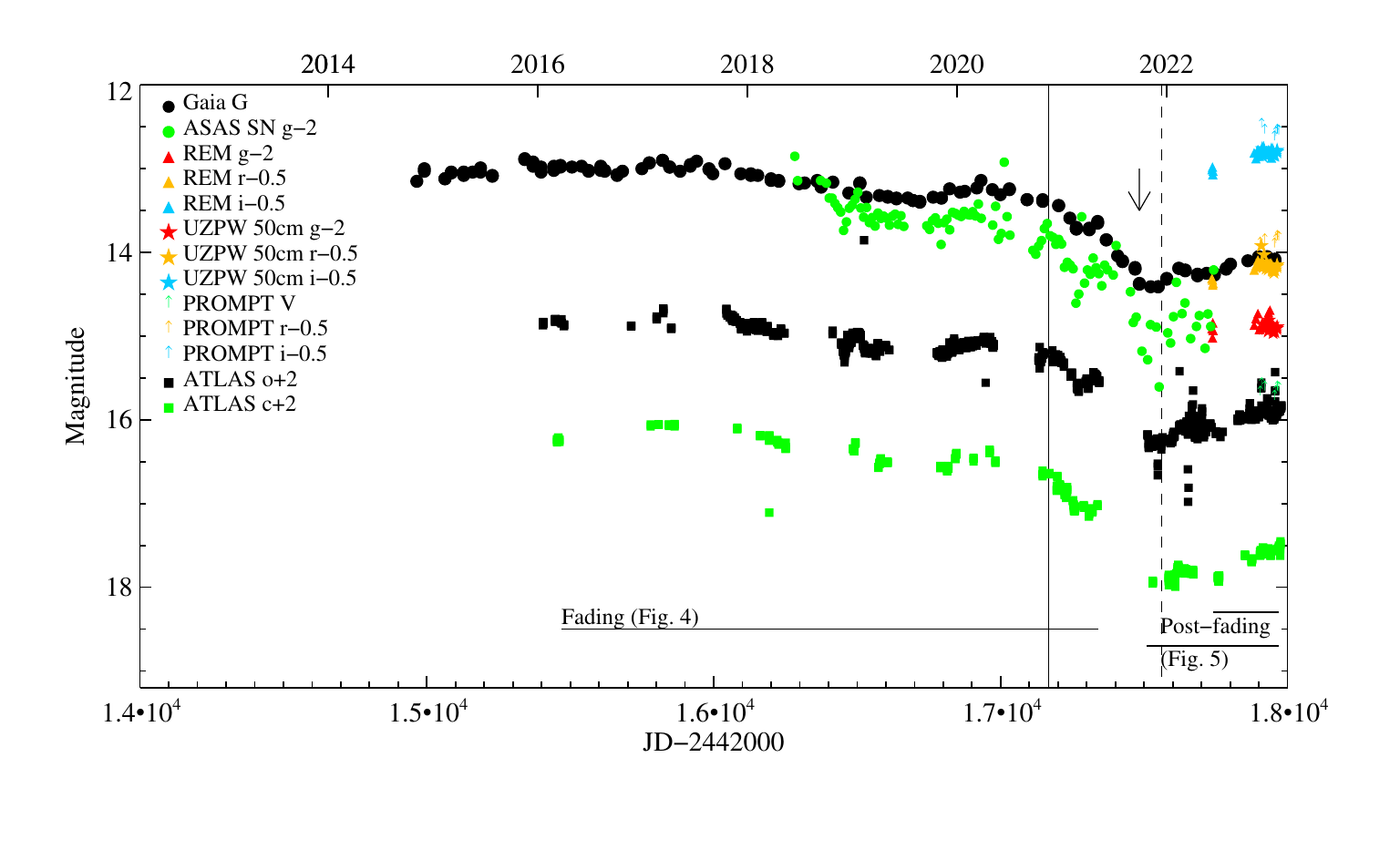}
\caption{Light curve of Gaia21elv in \textit{Gaia} G (black), \textit{WISE} $W1$ (blue) and $W2$ (grey) bands, and in $g$ band from the ASAS-SN (green). The solid vertical line shows the epoch of the Gemini/IGRINS spectrum, the dashed vertical line shows the epoch of the VLT/X-SHOOTER, and the arrow shows the epoch of the \textit{Gaia} alert. The ranges covered by the colour-magnitude diagrams in Figures \ref{fig:colormag_fading} and \ref{fig:colormag_followup} during and after the fading phase, respectively, are also indicated.}
\label{fig:lightcurve}
\end{figure*}

The results are shown in Fig. \ref{fig:lightcurve} and are summarized in Tables \ref{tab:phot1} and \ref{tab:phot2}. 

\subsection{Infrared photometry}

In 2022 June, we obtained infrared photometric observations with the REM, using the infrared imaging camera, REMIR. The reduction of the $JHK$ images, performed with our own IDL routines, included the construction and subtraction of a sky image, and flat-fielding. We extracted the instrumental magnitudes for the target as well as for all good-quality 2MASS stars (i.e. with a 2MASS photometric quality flag of AAA) in the field in an aperture with a radius of $\sim3\farcs7$. 
No extended nebulosity is visible around the source on the 2MASS images.
The final step was the determination of an average constant calibration factor between the instrumental and the 2MASS magnitudes of typically 30--50 stars, and this offset was applied to the target observations. The results can be found in Table \ref{tab:phot1}.

REMIR was used again between October 2022 and January 2023 for $J$-band imaging.
Each image came from the five single images jittered along a circle thanks to a dithering wedge from which a median sky was derived. Every image was then sky-subtracted with the median sky. Subsequently, the five images were re-aligned and averaged into a single $J$ band exposure. Calibrated images were then uploaded to the BHTOM service, reduced and matched to 2MASS $J$ band as described above for the optical data.

We used mid-infrared photometry from the Wide-field Infrared Survey Explorer (\textit{WISE}) and \textit{NEOWISE} surveys from the NASA/IPAC Infrared Science Archive. \textit{NEOWISE} observes the full sky on average twice per year with multiple exposures per epoch. For a comparison with the photometry from other instruments, we computed the average of multiple exposures of a single epoch.
\textit{NEOWISE} $W1$ and $W2$ photometry is known to display a photometric bias for saturated sources. We corrected for this bias using the correction curves given in the Explanatory Supplement to the \textit{NEOWISE} Data Release Products.
We derived the average of the uncertainties of the single exposures (err1). We also calculated the standard deviation of the points we averaged per season (err2). For the error of the data points averaged per epoch we used the maximum of err1 and err2.		

\subsection{Spectroscopy}

We obtained high-resolution (R$\sim$45,000) NIR spectra of Gaia21elv on 2020 November 14 (Program ID: GS-2020B-Q-218, PI: S. Park) using the Immersion GRating INfrared Spectrograph \citep[IGRINS;][]{yuk2010, park2014, mace2016} of Gemini South, in the $H$ and $K$ bands.
The spectrum was obtained with a slit size of 0.34\arcsec~$\times$~5\arcsec.
Gaia21elv was observed with two sets of ABBA nodding observations to subtract the sky background better. The total exposure time of Gaia21elv was 192~sec with 24~sec exposure of each frame. 
The data were reduced using the IGRINS pipeline \citep{lee2017} for flat-fielding, sky subtraction, correcting the distortion of the dispersion direction, wavelength calibration, and combining the spectra. 
In order to correct for telluric absorption features, a nearby A0 telluric standard star (HIP~21514) was observed right before the target. Then, the telluric correction and flux calibration were applied as done in \citet{park2018}.
Finally, barycentric velocity correction using barycorrpy \citep{kanodia2018} was applied ($V_{\rm{bary}}$ = 16.715\,km s$^{-1}$). 

A spectrum using the X-SHOOTER instrument of the Very Large Telescope (VLT) at ESO's Paranal Observatory in Chile \citep{Vernet2011} was taken on 2021 December 12 (Program ID: 108.23M6, PI: Z. Nagy). X-SHOOTER simultaneously covers a wavelength range from 300 nm to 2480 nm, and the spectra are divided into three arms, the ultraviolet (UVB, 300 -- 550 nm), the visible (VIS, 500 -- 1020 nm), and the near-infrared (NIR, 1000 -- 2480 nm). The observations were performed with the narrow slits of 1$''$, 0.9$''$, and 0.4$''$ in the UVB, VIS, and NIR respectively, leading to spectral resolution of R $\sim$ 5400, 8900, and 11600, respectively. The exposure time was 1800 s in each of the three arms. We obtained additional exposures with the 5$''$ slits, which resulted in data without slit losses, which we used for the correct flux calibration of the spectra obtained with the narrower slits. The ABBAAB nodding pattern was used.
The observations were processed with the official ESO pipeline. Telluric correction was performed using ESO's Molecfit program \citep{Kausch2015,Smette2015} running in the same EsoReflex environment \citep{Freudling2013}.

\begin{figure}
\centering
\includegraphics[width=6.5cm, trim={0cm 0.3cm 0cm 0cm},clip]{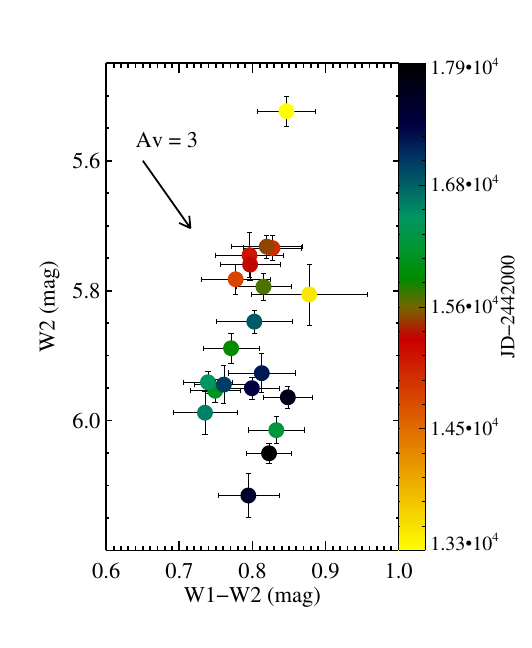}
\caption{Colour-magnitude diagram based on \textit{WISE} $W1$ and $W2$ data.}
\label{fig:colormag}
\end{figure}

\section{Results}
\label{sect_results}

\subsection{Light and colour variations}

Figure \ref{fig:lightcurve} shows the optical and infrared light curves of Gaia21elv, including archival data from 1977 (\citealp{ContrerasPena2019} and references therein), the All-Sky Automated Survey for Supernovae (ASAS-SN, \citealp{Shappee2014}, \citealp{Kochanek2017}), and the Asteroid Terrestrial-impact Last Alert System (ATLAS, \citealp{Tonry2018}, \citealp{Smith2020}, \citealp{Heinze2018}) survey downloaded from the ATLAS Forced Photometry web service \citep{Shingles2021}. 
Based on these data, the eruption occurred around between 1991 and 1996.
The amplitude of the brightening was 4-4.5 mag from a quiescent 16.5-17 mag to around 12 mag in the $R$-band. A slow fading of the source is already seen after 2010 based on data points from \citet{ContrerasPena2019} (collected from the AAVSO Photometric All Sky Survey (APASS) DR9 \citep{henden2015}, the VST Photometric Halpha Survey (VPHAS$+$) DR2 \citep{Drew2014}, the Bochum Galactic disc survey \citep{Hackstein2015}), and the \textit{Gaia} $G$-band light curve. 

In 2021, the source started a more rapid fading, and had a \textit{Gaia} alert in 2021 October due to its 1.2 mag fading in 18 months. After the \textit{Gaia} alert, a temporary brightening by about 0.2 mag was seen in early 2022, and after that, the source stayed at the same brightness for several months, around 14.25 mag in \textit{Gaia} $G$-band. Between 2022 July and November, the source brightened again, by about 0.3 mag as is seen in the lower panel of Fig.~\ref{fig:lightcurve}. A slow long-term fading is also seen in the \textit{WISE} data points.

Figure \ref{fig:colormag} shows a colour-magnitude diagram based on the \textit{WISE} $W1$ and $W2$ bands.
As the changes are mostly grey, extinction can be excluded as the physical mechanism between the flux changes observed at the \textit{WISE} wavelengths.

\begin{figure}
\centering
\includegraphics[width=9cm, trim={0cm 0 0 2cm},clip]{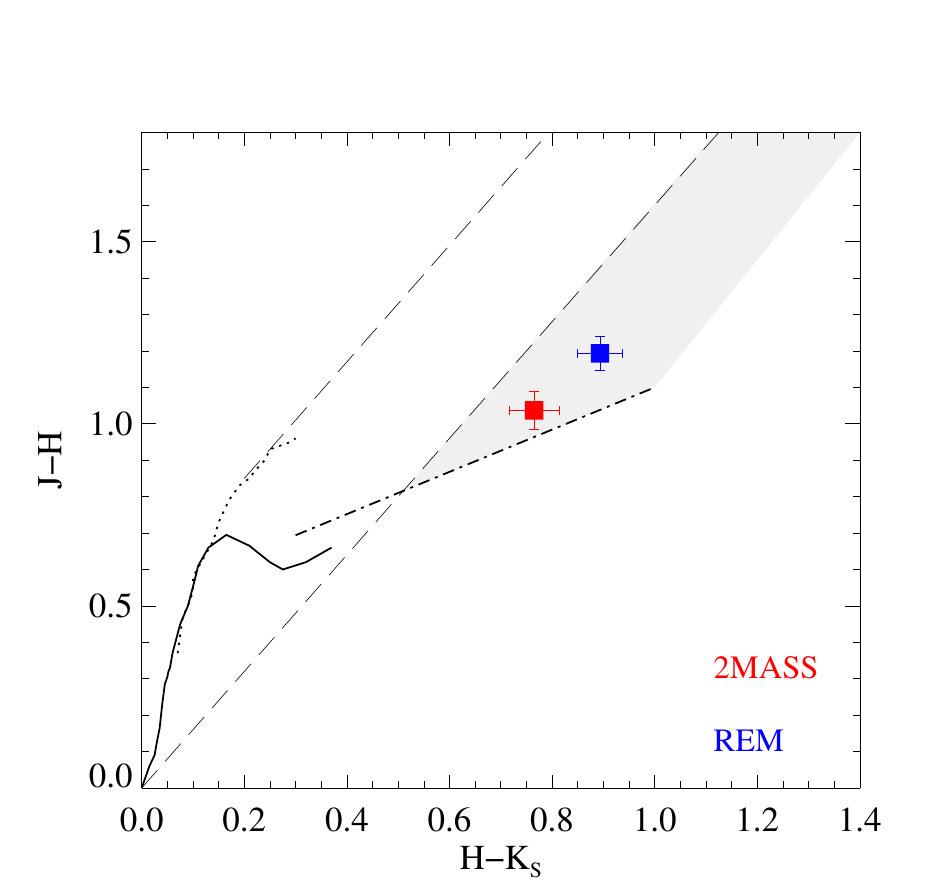}
\caption{($J-H$) versus ($H-K_S$) colour–colour diagram for the bright state (2MASS data point from 1999 February) and during the fading (REM data point from 2022 June). 
The solid curve shows the colours of the zero-age main-sequence, and the dotted line
represents the giant branch \citep{BessellBrett1988}. The long-dashed lines
delimit the area occupied by the reddened normal stars \citep{Cardelli1989}.
The dash–dotted line is the locus of unreddened CTTS \citep{Meyer1997} and the grey shaded band borders the area of the reddened $K_S$-excess stars.}
\label{fig:infra_color}
\end{figure}

Figure \ref{fig:infra_color} shows the $J-H$ vs $H-K_s$ diagram for the bright state (2MASS data point from 1999 February) and for the faint state (REM data point from 2022 June). 
The difference between the two data points in this diagram ($\Delta J \sim 0.61$ mag, $\Delta (J-H) \sim 0.16$ mag, $\Delta (H-K_s) \sim 0.13$ mag) may be consistent with the reddening of the source between 1999 and 2022. In this case, the colour change implies a visual extinction increase by $A_V \sim 2$ mag. 
However, the colour change in the $J-H$ vs $H-K_s$ diagram can also be caused by accretion. Eruptive young stars in the $J-H$ vs $H-K_s$ plot usually move toward or away from the main sequence (e.g. \citealp{SzegediElek2020}).

\begin{figure}
\centering
\includegraphics[width=6.8cm, trim={0cm 0.6cm 0cm 0cm},clip]{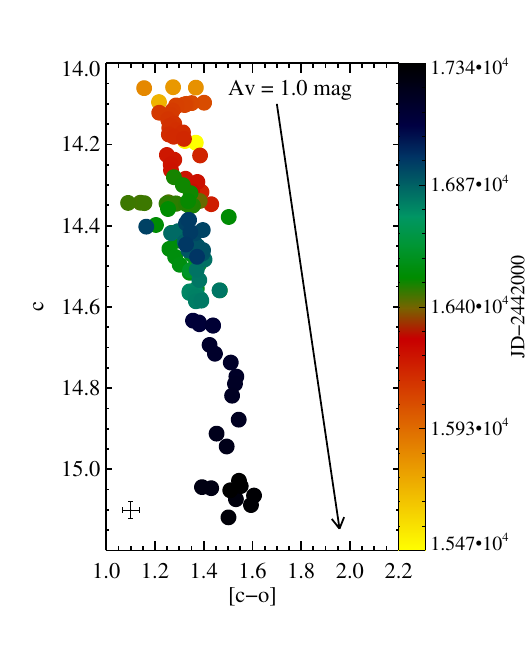}
\caption{Colour-magnitude diagram based on $o$ and $c$ magnitudes from the ATLAS survey during the fading of Gaia21elv. The typical error bar is shown in the lower left corner.}
\label{fig:colormag_fading}
\end{figure}

Figure \ref{fig:colormag_fading} shows a colour-magnitude diagram during the fading, as shown in Fig. \ref{fig:lightcurve} based on the $o$ and $c$ band magnitudes from the ATLAS survey. There is an indication of a long-term increasing trend of the extinction. Since the period of the quick fading in 2021 is not sampled well by these data points (as seen in Fig. \ref{fig:lightcurve}), it is not clear based on them, whether the increasing extinction also applies for this period.
 
\begin{figure*}
\centering
\includegraphics[width=5.7cm, trim={0cm 0.3cm 0cm 0cm},clip]{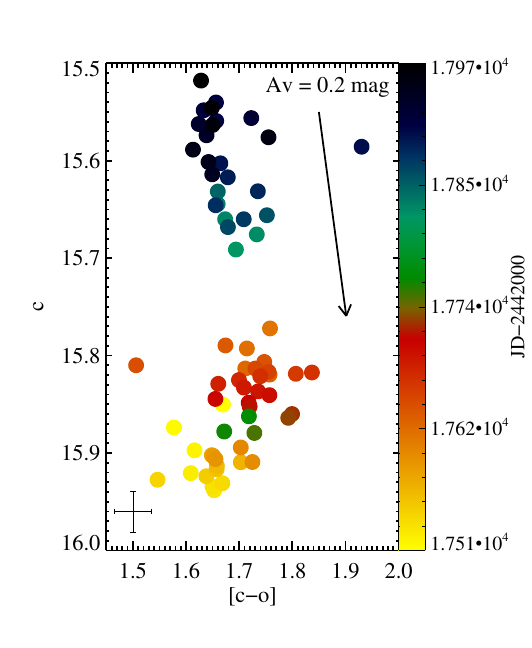}
\includegraphics[width=5.7cm, trim={0cm 0.3cm 0cm 0cm},clip]{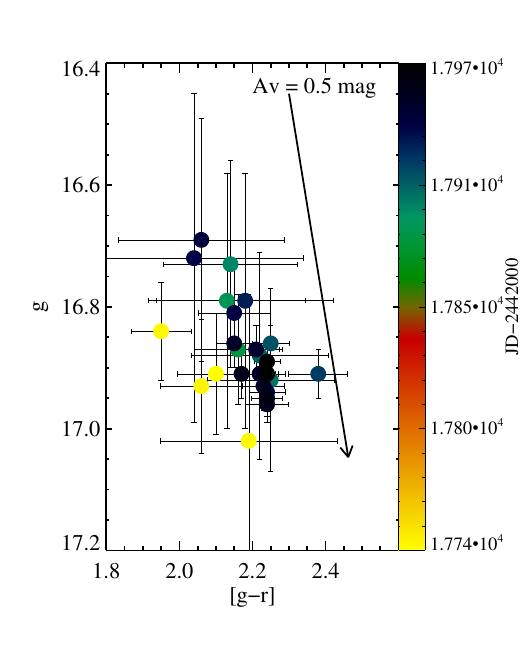}
\includegraphics[width=5.7cm, trim={0cm 0.3cm 0cm 0cm},clip]{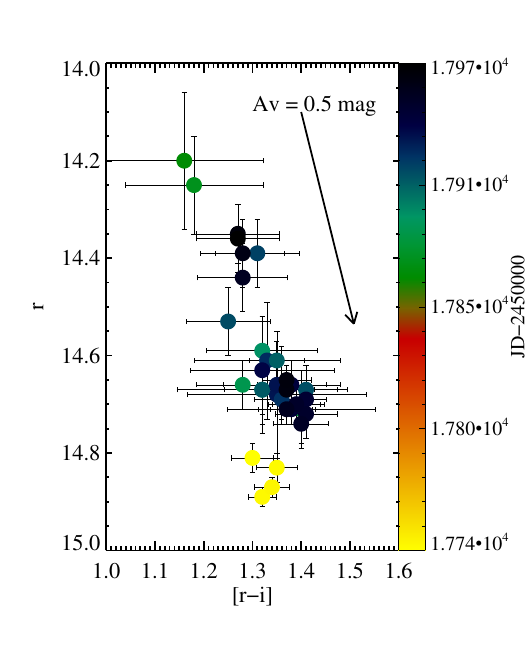}
\caption{
\textit{Left panel:} Colour-magnitude diagram based on $o$ and $c$ magnitudes from the ATLAS survey after the fading of Gaia21elv. The typical error bar is shown in the lower left corner.
\textit{Middle and right panels:} Colour-magnitude diagrams based on follow-up photometry shown in Tables \ref{tab:phot1} and \ref{tab:phot2}.
}
\label{fig:colormag_followup}
\end{figure*}

 Figure \ref{fig:colormag_followup} shows colour-magnitude diagrams after the fading of the source, based on the $o$ and $c$ band magnitudes from the ATLAS survey, $g-r$ versus $g$ and $r-i$ versus $r$ colour-magnitude diagrams based on our follow-up observations between 2022 June and 2023 January. The periods covered by these figures are also indicated in Fig. \ref{fig:lightcurve}. These colour-magnitude diagrams show extinction-related variations between 2022 June and 2023 January. The colour-magnitude diagram based on the ATLAS $o$ and $c$ band also includes data points from a period between 2021 October and 2022 May. These data points do not show an extinction-related trend, indicating, that mechanisms other than the extinction may also play a role in this post-fading phase.

Based on the colour variations alone, it is not possible to make a conclusion on the origin of the brightness variations of Gaia21elv. The $o$ and $c$ band data from the ATLAS survey as well as the $g-r$ versus $g$ and $r-i$ versus $r$ colour-magnitude diagrams suggest extinction-related brightness variations both during the fading and the brightening. Such extinction-related variations are not seen in the \textit{WISE} colour-magnitude diagrams, whereas the $J-H$ vs $H-K_s$ diagram can be interpreted both as a result of extinction and accretion. Therefore, we do not make a conclusion on the origin of the brightness variations based on the colour variations, and will further investigate it in Sect. \ref{sect:sed}.

\subsection{Reddening and spectral features}

\begin{figure*}
\centering
\includegraphics[width=16cm]{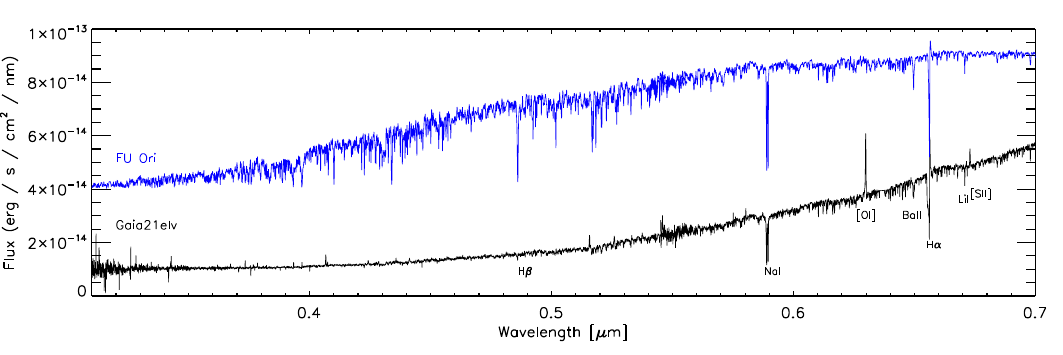}
\includegraphics[width=16cm]{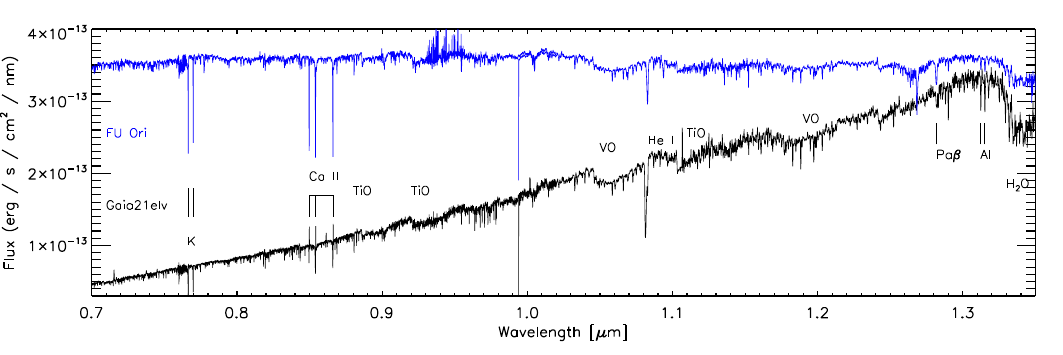}
\includegraphics[width=16cm]{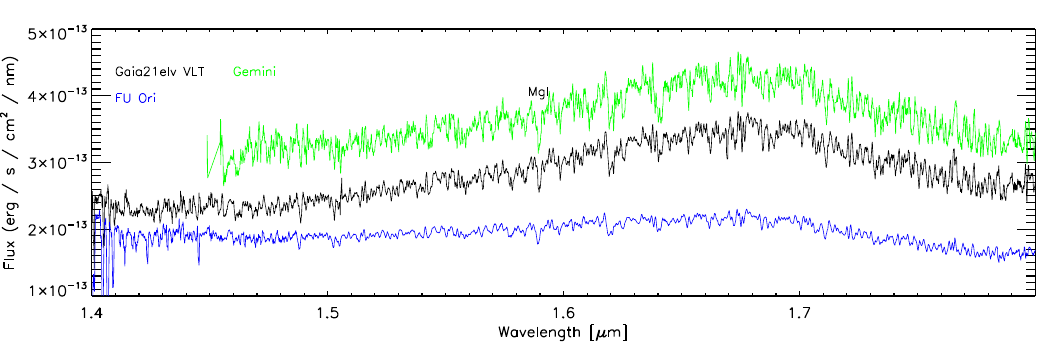}
\includegraphics[width=16cm]{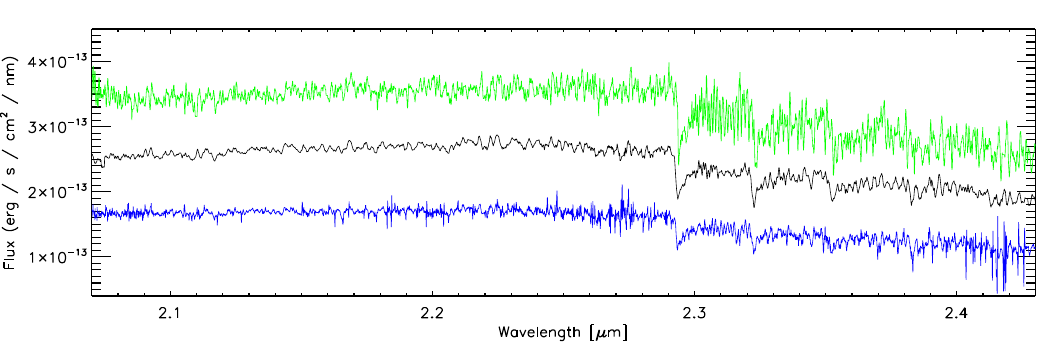}
\caption{Optical and NIR spectra of Gaia21elv taken with VLT/X-SHOOTER and Gemini South/IGRINS in comparison with those of FU Ori (taken also with VLT/X-SHOOTER, ESO archival data from program 094.C-0233). Arbitrary scaling factors were applied for a better comparison of the spectra. The Gemini South/IGRINS spectrum was smoothed for a better comparison.}
\label{fig:full_spectra}
\end{figure*}

Figure \ref{fig:full_spectra} shows the spectra taken at the two epochs in optical and NIR using Gemini South/IGRINS and VLT/X-SHOOTER and their comparison to the VLT/X-SHOOTER spectrum of FU Ori.

Following the method of \citet{ConnelleyReipurth2018}, we used the X-SHOOTER spectrum to estimate the visual extinction toward the source by comparing it to the spectrum of FU Ori, which has a low and well known extinction ($A_V = $1.7$\pm$0.1 mag; e.g. \citealp{Siwak2018}, \citealp{Lykou2022}).
We dereddened the spectrum of Gaia21elv with increasing $A_V$ until it matched the scaled, flux calibrated spectrum of FU Ori.
The resulting $\Delta A_V$ is $\sim$4 mag, which suggests $A_V \sim 5.7$ mag for Gaia21elv in its faint state.
 
\begin{figure}
\centering
\includegraphics[width=9cm, trim={0cm 0cm 0 0cm},clip]{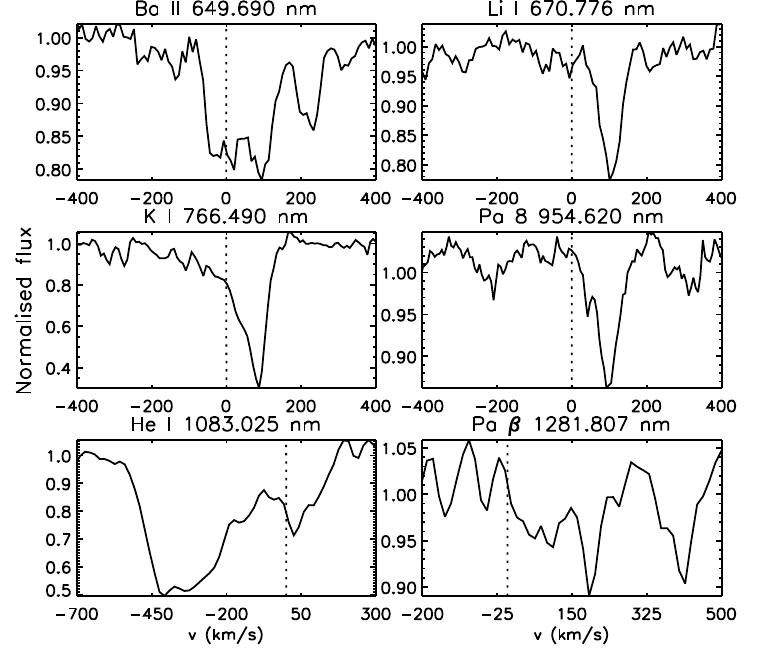}
\caption{Examples of absorption lines detected toward Gaia21elv.}
\label{fig:abs_lines}
\end{figure}

\begin{figure}
\centering
\includegraphics[width=9cm, trim={0cm 0cm 0 0.5cm},clip]{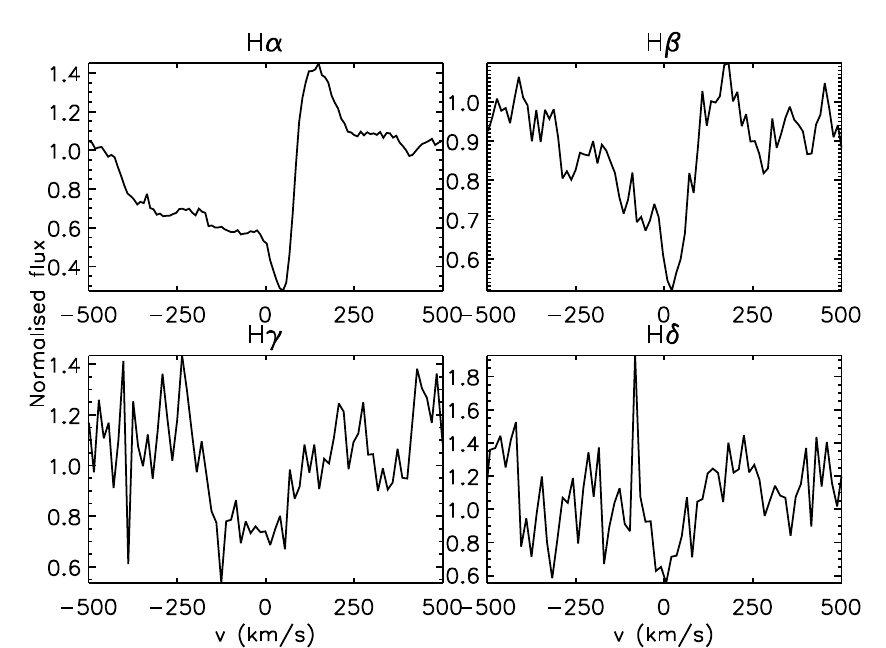}
\caption{Hydrogen Balmer lines in the VLT/X-SHOOTER spectrum of Gaia21elv.}
\label{fig:h_balmer}
\end{figure}

\begin{figure}
\centering
\includegraphics[width=8cm]{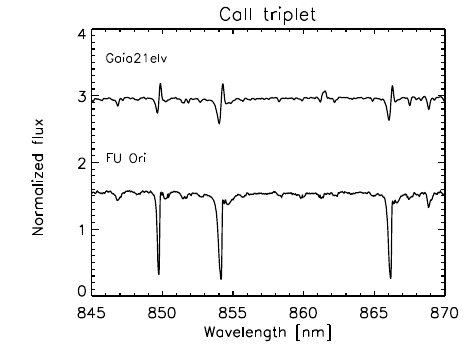}
\caption{Ca\,II triplet lines observed for Gaia21elv using VLT/X-SHOOTER, compared to those observed using VLT/X-SHOOTER for FU Ori.}
\label{fig:ca_triplet}
\end{figure}

\begin{figure}
\centering
\includegraphics[width=9cm, trim={0cm 0cm 0 0cm},clip]{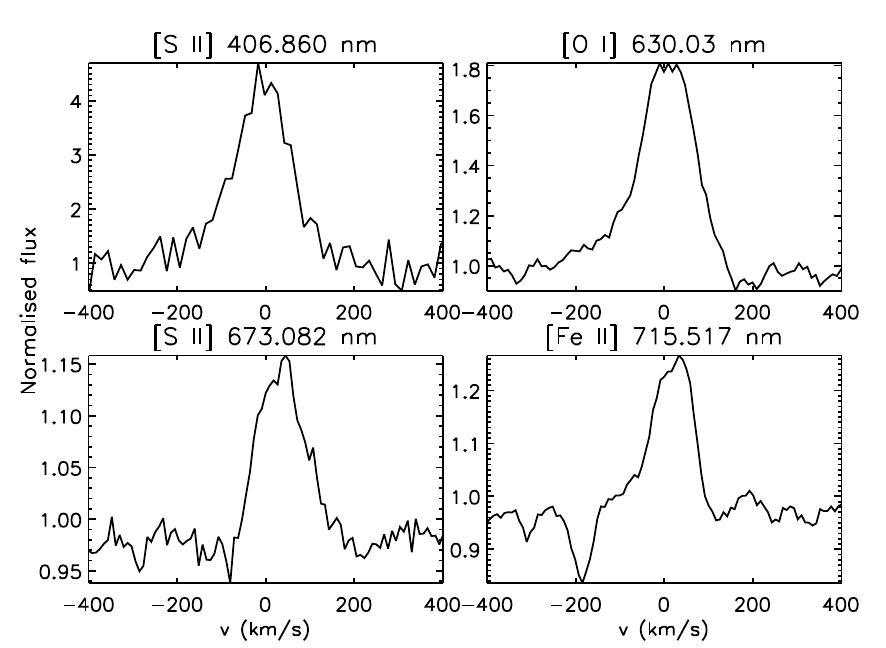}
\caption{Forbidden lines in emission detected toward Gaia21elv.}
\label{fig:forbidden_lines}
\end{figure}

\begin{figure}
\centering
\includegraphics[width=9cm, trim={0cm 0cm 0 0cm},clip]{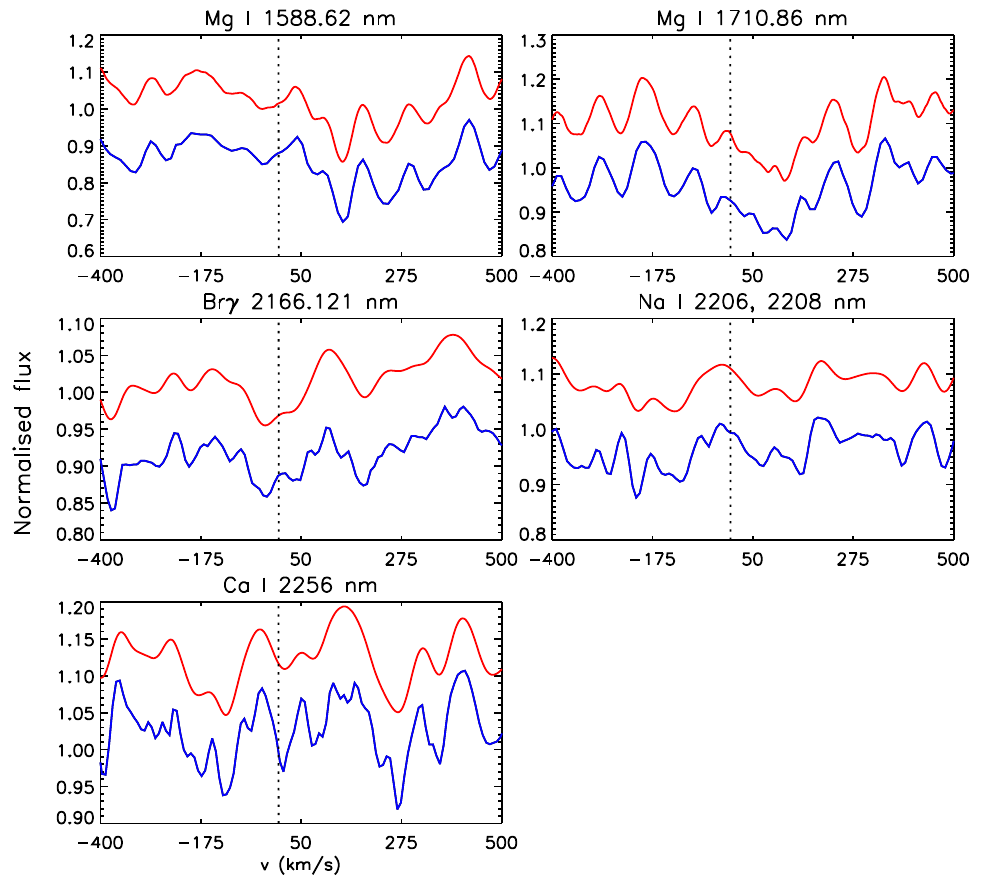}
\caption{Comparison of lines detected at both epochs using Gemini South/IGRINS (red) and VLT/X-SHOOTER (blue). Arbitrary scaling factors were applied for a better comparison of the spectra.}
\label{fig:lines_comparison}
\end{figure}

Table \ref{tab:lines} lists the lines we identified in the VLT/X-SHOOTER spectrum of Gaia21elv.
Most detected lines are seen in absorption, such as \ion{Ba}{ii}, \ion{Li}{i}, Na\,D, \ion{K}{i}, \ion{Al}{i}, \ion{He}{i}, Pa$\beta$, and \ion{Mg}{i} (Fig. \ref{fig:abs_lines}). Some of these absorption lines show two (or more) components, such as the \ion{Ba}{ii}, \ion{He}{i}, and Pa$\beta$ lines. 
Some lines show a P Cygni profile, such as H$\alpha$ and H$\beta$ (Fig. \ref{fig:h_balmer}) and the \ion{Ca}{ii} triplet (Fig. \ref{fig:ca_triplet}).
Forbidden lines of [\ion{O}{i}], [\ion{Fe}{ii}], and [\ion{S}{ii}]~were detected in emission (Fig. \ref{fig:forbidden_lines}). These lines may indicate the presence of a jet associated with Gaia21elv, similarly to what was seen for the classical FUor V1057 Cyg (e.g. \citealp{Szabo2021}). Forbidden emission lines in young stars were also suggested to trace disk winds (\citealp{PaatzCamenzind1996}, \citealp{IguchiItoh2016}, \citealp{Ballabio2020}).
The $H$ and $K$-band spectra were observed at two different epochs: in 2020 November, just before the rapid fading of the source (Gemini South/IGRINS) and in 2021 December, soon after the \textit{Gaia} alert reporting the fading (VLT/X-SHOOTER). These spectra display very similar features (Fig. \ref{fig:full_spectra}), including a triangular shaped $H$-band continuum and the CO-bandhead features in absorption, both typical FUor signatures. 
Fig. \ref{fig:lines_comparison} shows lines detected at both epochs, such as \ion{Mg}{i}, Br$\gamma$, \ion{Na}{i}, and \ion{Ca}{i}. The line profiles did not change significantly between the two epochs.

\begin{table*}
\centering
\caption{Lines detected in the X-SHOOTER spectrum of Gaia21elv. The FWHM values were derived using a Gaussian fitting, and are not provided for line profiles, which cannot be fitted by a Gaussian. For lines with multiple components, we provide the parameters of the one with the highest intensity.}
\label{tab:lines}
\begin{tabular}{cccccc}
\hline
Species& Lab. $\lambda$& Obs. $\lambda$& EW&   FWHM& Note\\ 
       &           [nm]&           [nm]& [nm]& [nm]&     \\
\hline
$[$\ion{S}{ii}$]$&  406.860& 406.702& $-$1.01$\pm$0.05& 0.20$\pm$0.03& emission\\
H$\delta$&   410.171& 410.038&    0.05$\pm$0.01& 0.10$\pm$0.01& absorption\\
H$\gamma$&   434.047& 433.850&    0.13$\pm$0.03& 0.32$\pm$0.01& absorption\\
H$\beta$&    486.129& 485.997&    0.13$\pm$0.01&           ...& P Cygni absorption\\
H$\beta$&    486.129& 486.251& $-$0.04$\pm$0.01& 0.16$\pm$0.02& P Cygni emission\\
Na\,D&        588.995& 588.898&    0.30$\pm$0.01& 0.27$\pm$0.01& absorption\\
Na\,D&        589.592& 589.508&    0.25$\pm$0.01& 0.24$\pm$0.02& absorption\\
$[$\ion{O}{i}$]$&   630.030& 629.801& $-$0.28$\pm$0.02& 0.29$\pm$0.01& emission\\
\ion{Ba}{ii}&       649.690& 649.515&    0.08$\pm$0.01& 0.38$\pm$0.01& absorption\\
H$\alpha$&   656.282& 656.155&    0.42$\pm$0.01&           ...& P Cygni absorption\\
H$\alpha$&   656.282& 656.377& $-$0.09$\pm$0.01& 0.18$\pm$0.02& P Cygni emission\\
\ion{Li}{i}&        670.776& 670.785&    0.03$\pm$0.01& 0.13$\pm$0.01& absorption\\
$[$\ion{S}{ii}$]$&  673.082& 672.960& $-$0.05$\pm$0.01& 0.28$\pm$0.01& emission\\
$[$\ion{Fe}{ii}$]$& 715.517& 715.364& $-$0.08$\pm$0.01& 0.24$\pm$0.01& emission\\
\ion{K}{i}&         766.490& 766.457&    0.15$\pm$0.01& 0.20$\pm$0.02& absorption\\
\ion{K}{i}&         769.896& 769.851&    0.10$\pm$0.01& 0.20$\pm$0.02& absorption\\
\ion{Ca}{ii}&       849.802& 849.647&    0.05$\pm$0.01& 0.19$\pm$0.02& P Cygni absorption\\
\ion{Ca}{ii}&       849.802& 849.854& $-$0.04$\pm$0.01& 0.12$\pm$0.01& P Cygni emission\\
\ion{Ca}{ii}&       854.209& 854.028&    0.10$\pm$0.01& 0.28$\pm$0.01& P Cygni absorption\\
\ion{Ca}{ii}&       854.209& 854.276& $-$0.06$\pm$0.01& 0.16$\pm$0.01& P Cygni emission\\
\ion{Ca}{ii}&       866.214& 866.043&    0.09$\pm$0.01& 0.23$\pm$0.01& P Cygni absorption\\
\ion{Ca}{ii}&       866.214& 866.283& $-$0.04$\pm$0.01& 0.14$\pm$0.01& P Cygni emission\\
Pa8&                954.620& 954.607&    0.04$\pm$0.01& 0.21$\pm$0.02& absorption\\
\ion{He}{i}&       1083.025& 1081.46&    0.60$\pm$0.10& 1.20$\pm$0.20& absorption, two components\\
Pa$\beta$&         1281.807& 1281.819&             ...& ...& absorption, two components\\
\ion{Al}{i}&       1312.342& 1312.382&   0.06$\pm$0.01& 0.18$\pm$0.01& absorption\\
\ion{Al}{i}&       1315.075& 1315.107&   0.07$\pm$0.01& 0.22$\pm$0.01& absorption\\
\hline
\end{tabular}
\end{table*}

\begin{figure}
\centering
\includegraphics[width=9cm]{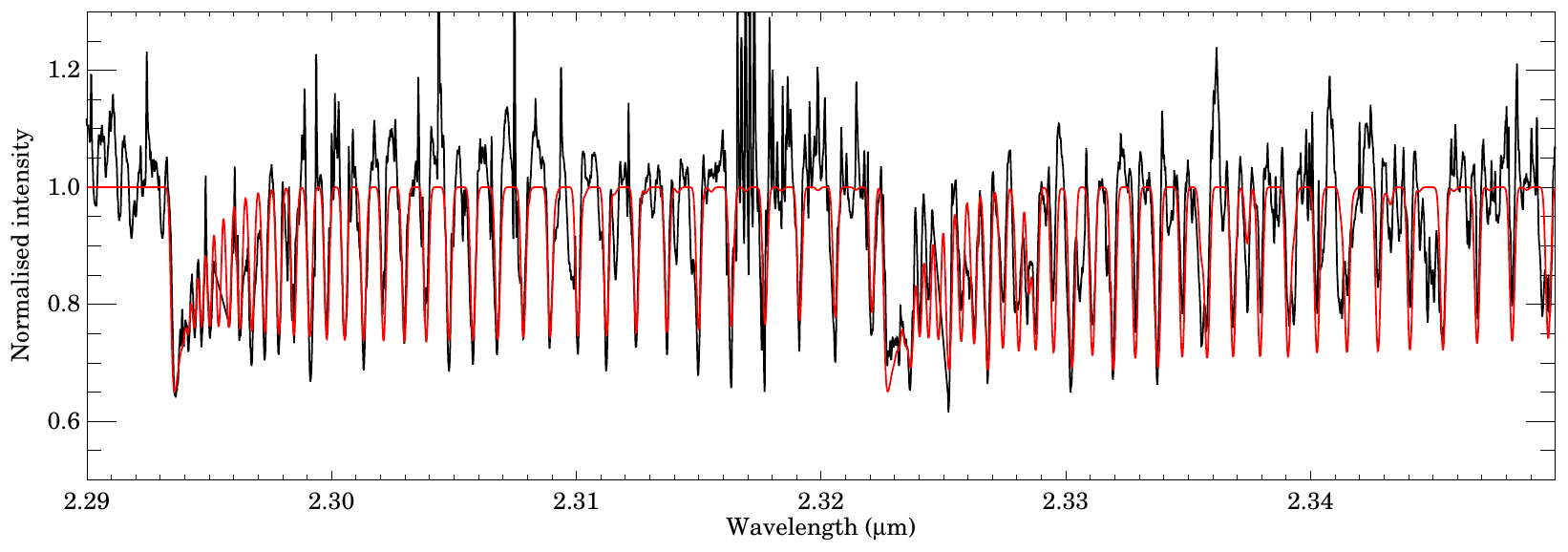}
\includegraphics[width=9cm]{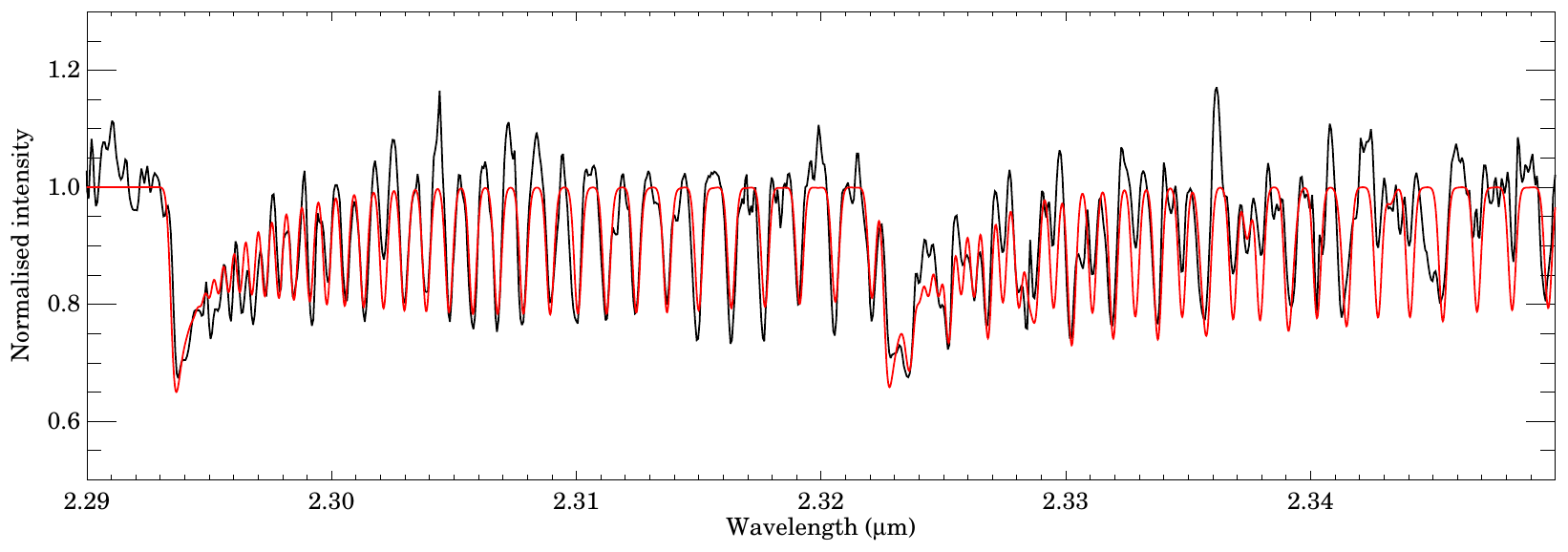}
\caption{CO overtone features of Gaia21elv shown in black observed using Gemini South/IGRINS (top panel) and VLT/X-SHOOTER (bottom panel). The best fit models are overplotted in red.}
\label{fig:co_fit}
\end{figure}

To interpret the CO bandhead features observed at the two epochs, we used an isothermal slab model to find a best-fitting CO column density and excitation temperature of the absorbing material, similarly to \citet{Kospal2011} and \citet{Park2021}. We found the best-fitting CO column density to be $\sim$10$^{22}$ cm$^{-2}$, and a best-fitting excitation temperature of $2800\pm100$~K at the first epoch (Gemini South/IGRINS) and $2300\pm100$~K at the later epoch (VLT/X-SHOOTER). The results are shown in Figure \ref{fig:co_fit}.
In Sect. \ref{gaia21elv_fuor} we analyse the spectra in more detail and compare the observed features to those seen in FUors.

\subsection{Spectral Energy Distribution modeling}
\label{sect:sed}

\begin{figure}
\centering
\includegraphics[width=9.1cm, trim={0.3cm 0 0 0},clip]{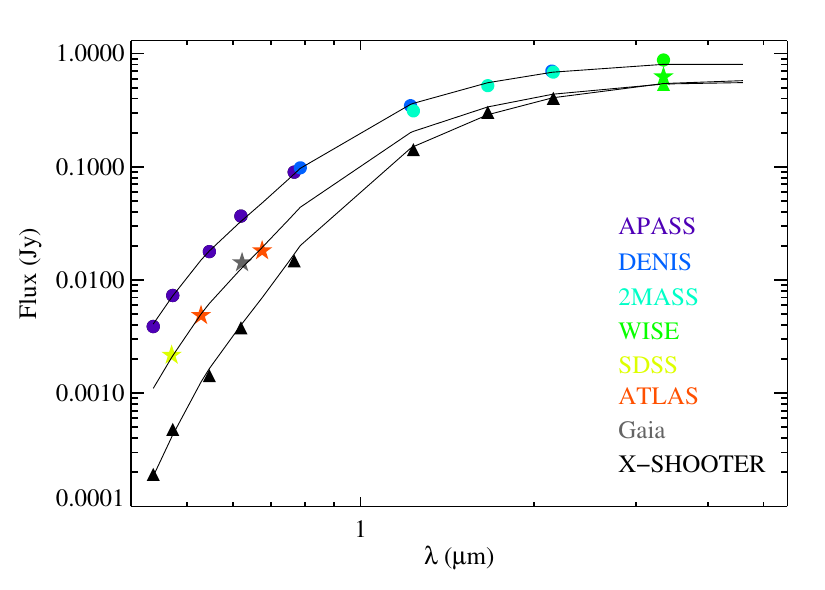}
\caption{The SED of Gaia21elv at the three modelled epochs. The SED at the brightest state based on archival data is shown with circles. The SED close to the epoch of the Gemini/IGRINS spectrum is shown with asterisks. The SED at the epoch of the VLT/X-SHOOTER spectrum representing the faint state is shown with triangles.
Solid curves show the results of the accretion disc models for the individual epochs.}
\label{fig:sed}
\end{figure}

In the following, we analyse the Spectral Energy Distribution (SED) of Gaia21elv at three different epochs. 
To create a SED for the state of the maximum brightness, we used archival data from APASS9 \citep{henden2015}, DENIS \citep{Epchtein1994}, 2MASS \citep{cutri2003}, and the ALLWISE \citep{wise_ref} catalogues. 
A comparison of the DENIS $I$-band flux from 1996 December with the APASS9 $i'$-band flux from 2010 December shows that the brightness of the star did not change significantly between these dates, thus the fact that the used archival data correspond to different epochs is not expected to affect the modeling of the SED in the bright state.
In addition to the epoch of the bright state, we compiled an SED for 2020
Oct--Nov that is very close to the epoch of the Gemini/IGRINS spectrum, and as such, it is just before the fast fading phase of the source. We used the available ASAS-SN $g$, \textit{Gaia} $G$ and \textit{WISE} $W1$ data, as well as photometry in the cyan and orange bands of the ATLAS survey for this epoch.
The third epoch we considered is the epoch of the VLT/X-SHOOTER spectrum in 2021 December, as it represents the faint state at the end of the fast fading of the source. We obtained synthetic photometry in the APASS9 and 2MASS bands from the X-SHOOTER spectrum, and also used the \textit{NEOWISE} $W1$ data point closest to this epoch. 
The three SEDs are shown in Fig. \ref{fig:sed}.

As we will discuss in Sec.~\ref{sect_discussion}, the properties of Gaia21elv resemble those of FU\,Orionis-type stars. In these objects the circumstellar matter is expected to form an accretion disc \citep{HartmannKenyon1996}. 
To estimate the properties of the accretion disc in Gaia21elv at the three epochs, we modelled the SEDs using a steady, optically thick and geometrically thin viscous accretion disc, whose mass accretion rate is constant in the radial direction.
This method was successfully applied to estimate the accretion rate in several eruptive YSOs including HBC~722 \citep{Kospal2016}, V582 Aur \citep{Abraham2018}, 2MASS 22352345 + 7517076 \citep{Kun2019}, Gaia18dvy \citep{SzegediElek2020}, V1057 Cyg \citep{Szabo2021}, and V1515 Cyg \citep{Szabo2022}. 
In this model, the temperature profile of the disc is defined based on \citet{HartmannKenyon1996} as:
\begin{equation} 
T (r) = \left[ \frac{3GM_\star \dot{M}}{8\pi R^3_\star \sigma} \left( 1 - \sqrt{\frac{R_\star}{r}} \right) \right]^{1/4},
\end{equation}
where $r$ is the distance from the star, $R_\star$ is the stellar radius, $M_\star$ is the stellar mass, $\dot{M}$ is the accretion rate, and $G,\sigma$ are the gravitational and Stefan-Boltzmann constants, respectively. 
The model SED was calculated by integrating black-body emission in concentric annuli between the inner disc radius and the outer disc radius. The resulting SED was then reddened by different A$_V$ values.

One of the input parameters of the model is the inclination, and as it is unknown for Gaia21elv, we used an intermediate value of 45$^\circ$. 
We assumed a distance of $910.9$ pc, as derived above from the \textit{Gaia} DR3 parallax and its zero-point correction.
There is a known degeneracy in the model between the inner disc radius and A$_V$. To break this degeneracy we adopted the A$_V$ value of $\sim$5.7 mag obtained from the X-SHOOTER spectrum in Sect.~\ref{sect_results}. This choice fixed the inner disc radius to $R_{\rm{in}} = 2 R_\odot$, 
a reasonable value, as it is the same as determined for FU Ori by \citet{Zhu2007}. 

\begin{figure}
\centering
\includegraphics[width=8.5cm, trim={0cm 0 -0.3cm 0},clip]{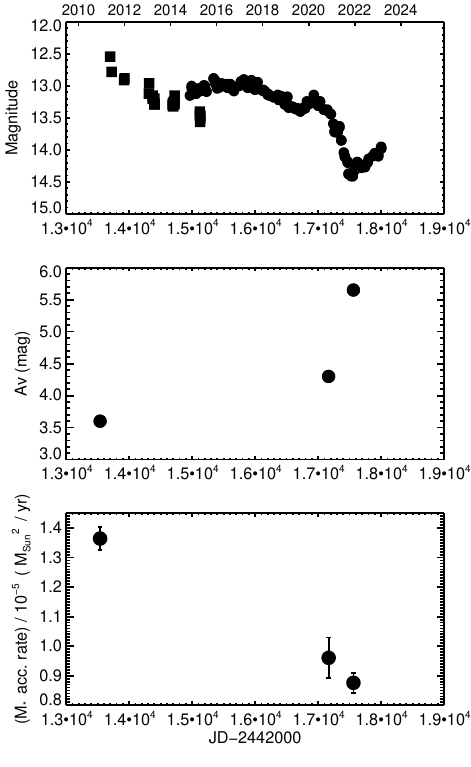}
\caption{Visual extinctions and the product of the stellar mass and accretion rate for the three epochs based on the accretion disc models. The long-term light curve using the \textit{Gaia} $G$ magnitudes and the data from \citet{ContrerasPena2019} are shown as a comparison.}
\label{fig:diskmodel_results}
\end{figure}

The remaining free parameters of the disc model are $M_\star \dot{M}$, A$_V$, and $R_{\rm{out}}$. Finding the best $M_\star \dot{M}$ and A$_V$ combinations was performed with $\chi^2$ minimization over a large grid in both the accretion rate and the extinction, by taking into account all flux values between 0.4 and 4.0 $\mu$m. The formal uncertainties
of the data points were set to a homogeneous 5\% of the measured flux values. We ran several models assuming different outer disc radii in the range between 0.2 and 2 au, and found that the WISE data points are reasonably well fitted with $R_{\rm{out}} = 1$ au, though this value is less constrained than the other two parameters. The best-fitting visual extinctions and products of the stellar mass and the accretion rate are plotted in Fig. \ref{fig:diskmodel_results}. Since the outcome of the model is the product $M_\star \dot{M}$, the true accretion rate depends on the stellar mass. However, FUors are typically low-mass objects \citep{HartmannKenyon1996}, thus our obtained values provide a good approximation to the accretion rate.

Considering the results for all three epochs, the three data points suggest that the accretion rate followed a monotonic decay in the last 15 years. Our models suggest a slight increase of the extinction toward the source from 3.6\,mag to 4.4\,mag between the maximum brightness and the Gemini epoch in 2020 November. Remarkably, the quick fading in 2021, corresponding to the \textit{Gaia} alert, was mostly caused by an increase in extinction. The accretion luminosity of the source also dropped in parallel to the accretion rate between the first and last epoch, from 106\,L$_\odot$ to 68\,L$_\odot$, although the absolute values depend on the unknown inclination angle, too.

\section{Discussion}
\label{sect_discussion}

\subsection{Classification of Gaia21elv as a FUor}
\label{gaia21elv_fuor}

To investigate, whether Gaia21elv is indeed a FUor, we used the criteria from \citet{ConnelleyReipurth2018}, which they list in their Table 3. In the following, we list these defining characteristics and check if Gaia21elv fulfills them.

\begin{itemize}

\item[-] The eruption is observed for each bona fide FUor, unlike for FUor-like and peculiar objects. This criterion is fulfilled for Gaia21elv. The date of the eruption can be constrained based on the light curve shown in \citet{ContrerasPena2019} in their figure B2, which includes data points from the literature starting from 1977 (Fig. \ref{fig:lightcurve}). 
The outburst of Gaia21elv based on the long term light curve occurred between 1991 and 1996. 

\item[-] Bona fide FUors have well defined CO absorption features. Strong CO absorption was also observed for Gaia21elv (Fig. \ref{fig:co_fit}) at both of our observing epochs.

\item[-] Water vapor bands can be identified in the NIR spectra of bona fide FUors, including the feature at 1.33 $\mu$m and the triangular shaped $H$-band continuum, which is due to water vapor bands on each end of the $H$-band (Fig. \ref{fig:full_spectra}). Gaia21elv shows these features at both epochs.

\item[-] Bona fide FUors show other molecular bands in their $J$-band spectra, such as those from vanadium oxide (at 1.05 $\mu$m and 1.19 $\mu$m) and titanium oxide (0.88, 0.92, and 1.11 $\mu$m). The X-SHOOTER spectrum of Gaia21elv shows all these molecular bands as wide absorption features (Fig. \ref{fig:full_spectra}).

\item[-] Another characteristic of FUors is their hydrogen lines, especially the Pa$\alpha$, $\beta$, $\gamma$, and $\delta$ lines, are in absorption, Br$\gamma$ line is very weak, with the rest of the Brackett series not observed.  
For Gaia21elv the Pa$\beta$ and Pa$\delta$ lines are indeed seen in absorption, however, the other two Paschen lines are not detected. It was not possible to detect the Pa$\alpha$ line due to the poor atmospheric transmission at its wavelength (1.87 $\mu$m). 
The Br$\gamma$ line shows a weak absorption, while the rest of the Brackett series is not detected, similarly to what was expected for FUors.

\item[-] FUors show very few, if any, emission lines, and even those are typically the emission components of P Cygni profiles. Gaia21elv shows a few P Cygni profiles in H$\alpha$, H$\beta$, and the Ca II triplet, and in addition to those, there are forbidden lines of [\ion{O}{i}], [\ion{Fe}{ii}], and [\ion{S}{ii}] in emission. The absorption lines and P Cygni profiles typically detected in the spectra of FUors are related to the disc, while the forbidden emission lines trace a jet or disk wind. 
Forbidden emission lines are not always detected in the spectra of known FUors, but were identified for a few examples, including the classical FUors V2494 Cyg (\citealp{ConnelleyReipurth2018} and references therein) and V1057 Cyg \citep{Szabo2021}, therefore, their detection does not rule out a classification as a bona fide FUor.

\item[-] FUors show weak absorption lines of \ion{Na}{i} (2.208 $\mu$m) and \ion{Ca}{i} (2.256 $\mu$m) \citep{ConnelleyReipurth2018}. As shown in Fig. \ref{fig:lines_comparison}, these lines are detected in the spectra of Gaia21elv at both epochs.

\item[-] Another spectroscopic signature of FUors is the \ion{He}{i} line at 1.083 $\mu$m, which is also present in the spectrum of Gaia21elv (Fig. \ref{fig:full_spectra}). The \ion{He}{i} line detected toward Gaia21elv is double-peaked, where the higher intensity component is largely blueshifted, detected at a velocity of around $-$400 km~s$^{-1}$, and the lower intensity component is seen at a velocity of around $+$25 km~s$^{-1}$ (Fig. \ref{fig:abs_lines}). 
Most bona fide FUors show blueshifted absorption lines, with a mean velocity of $-$350 km~s$^{-1}$ (see Fig. 4. in \citealp{ConnelleyReipurth2018}). 

Another characteristics of FUors is that their spectral type is wavelength-dependent \citep{HartmannKenyon1996}. 
To check whether this applies to Gaia21elv, we used the VLT/X-SHOOTER spectrum, and compared it to the synthetic stellar spectra calculated by \citet{Coelho2005} in the 300 nm to 1.8 $\mu$m wavelength range. These stellar templates are given for effective temperatures in the range between 3500~K and 6000~K in steps of 250~K. We compared the VLT/X-SHOOTER spectrum to these stellar templates at optical and at NIR wavelengths, separately. At optical wavelengths, the best match was found with the stellar template corresponding to an effective temperature of 5500$\pm$250~K, while at NIR wavelengths, the best fit corresponds to an effective temperature of 3750$\pm$250~K. This is consistent with the expectation for FUors, that the stellar type is wavelength-dependent.

Based on the above criteria from \citet{ConnelleyReipurth2018} as well as its wavelength-dependent spectral type, we conclude that Gaia21elv can be classified as a bona fide FUor. 
This classification is consistent with the high accretion luminosity of the source implied by our accretion disc modelling.

\end{itemize}

\subsection{On the recent fading of Gaia21elv}

Until now, no bona fide FUor is known to have completely ended its outburst. This is why it is important to monitor their brightness variations, and study their fading episodes.
A temporary fading of V346 Nor was reported by \citet{Kraus2016} and \citet{Kospal2020}, which was due to a decrease in the accretion rate, however, after the fading, the star brightened again to nearly reach its outburst brightness. 
Another eruptive young star, V899 Mon, which shows properties of both FUors and EXors, faded to quiescence for a little less than a year \citep{Ninan2015,Park2021}. However, this quiescent phase was followed by another outburst. 
In addition to their fading being temporary, neither V346 Nor, nor V899 Mon is a bona fide FUor.
The long-term fading of a classical FUor, V1515 Cyg was recently reported by \citet{Szabo2022}: its fading started around 2006 and is approximately consistent with an exponential decay with an e-folding time of 12 years. 
Another classical FUor, V733 Cep also shows long-term fading (Park et al., in prep.), which was found to be the result of a decrease in the accretion rate.

Brightness variations of young stars are only partly related to changes in the accretion rate \citep{Fischer2022}. The other main process is variable circumstellar extinction. To probe whether the fading of Gaia21elv was the result of a decrease of the accretion rate, we estimated the accretion rate by fitting the SEDs with an accretion disc model in Sec.~\ref{sect_results}.
The accretion rates derived for Gaia21elv are typical of FUors (\citealp{Fischer2022} and references therein). 
The accretion rate between the brightest and faintest states decreased by $\sim$36\%.
However, according to the accretion disc models fitted to the SEDs, the decreasing accretion rate was combined with increasing circumstellar extinction, especially between 2020 and 2022. 
It is most likely, that the increased circumstellar extinction dominated the rapid fading of the source that triggered the \textit{Gaia} Alerts system in 2021. 
After the \textit{Gaia} alert, the brightness of the source also started a slow increase, though it is still almost a magnitude fainter than in early 2020, before the start of this fading episode. 
The decrease found between the accretion rates at the brightest and faintest states indicates an e-folding time of about 25 years. 
Based on our results, the fading of Gaia21elv found by the \textit{Gaia} alert is likely a temporary event. Future photometric and spectroscopic monitoring of the source is important to provide more information on the evolution of its outburst. 

\section{Summary}
\label{sect_summary}

We analysed the photometry and spectroscopy of a young star exhibiting a long-term outburst and a recent fading alerted by the \textit{Gaia} Science Alerts system.

Optical and NIR spectra confirm that Gaia21elv is a bona fide FUor. This is the third FUor which was found based on the \textit{Gaia} alerts.
In addition to the classical FUor signatures, forbidden emission lines were detected, which are typically tracing a jet or disk winds.

Fitting the SEDs at the maximum brightness and and its faint state using an accretion disc model suggests a decrease in the accretion rate. However, fitting the SED at an epoch close to the onset of the quick fading in late 2020-2021 indicates that this episode was mostly caused by an increase of circumstellar extinction.

In the future, a photometric and spectroscopic monitoring of Gaia21elv is important to characterize its behavior after its fading episode.

\section*{Acknowledgements}

We thank the referee for comments which helped to improve our paper.

This project has received funding from the European Research Council (ERC) under the European Union's Horizon 2020 research and innovation programme under grant agreement No 716155 (SACCRED).

We acknowledge support from the ESA PRODEX contract nr. 4000132054.

G.M. and Z.N. were supported by the J\'anos Bolyai Research Scholarship of the Hungarian Academy of Sciences.

G.M. has received funding from the European Union’s Horizon 2020 research and innovation programme under grant agreement No. 101004141.

Zs.M.Sz. acknowledges funding from a St Leonards scholarship from the University of St Andrews. For the purpose of open access, the author has applied a Creative Commons Attribution (CC BY) licence to any Author Accepted Manuscript version arising.

E.F. and T.G. acknowledge financial support from the project PRIN-INAF 2019 "Spectroscopically Tracing the Disk Dispersal Evolution (STRADE)".

We acknowledge ESA Gaia, DPAC and the Photometric Science Alerts Team (http://gsaweb.ast.cam.ac.uk/alerts).

This work used the Immersion Grating Infrared Spectrometer (IGRINS) that was developed under a collaboration between the University of Texas at Austin and the Korea Astronomy and Space Science Institute (KASI) with the financial support of the Mt. Cuba Astronomical Foundation, of the US National Science Foundation under grants AST-1229522 and AST-1702267, of the McDonald Observatory of the University of Texas at Austin, of the Korean GMT Project of KASI, and Gemini Observatory.

This work was supported by K-GMT Science Program (PID: GS-2020B-Q-218) of Korea Astronomy and Space Science Institute (KASI).

Based on observations collected at the European Southern Observatory under ESO programme 108.23M6.

This work has made use of data from the Asteroid Terrestrial-impact Last Alert System (ATLAS) project. The Asteroid Terrestrial-impact Last Alert System (ATLAS) project is primarily funded to search for near earth asteroids through NASA grants NN12AR55G, 80NSSC18K0284, and 80NSSC18K1575; byproducts of the NEO search include images and catalogs from the survey area. This work was partially funded by Kepler/K2 grant J1944/80NSSC19K0112 and HST GO-15889, and STFC grants ST/T000198/1 and ST/S006109/1. The ATLAS science products have been made possible through the contributions of the University of Hawaii Institute for Astronomy, the Queen’s University Belfast, the Space Telescope Science Institute, the South African Astronomical Observatory, and The Millennium Institute of Astrophysics (MAS), Chile.

This project used data obtained via BHTOM (https://bhtom.space), which has received funding from the European Union's Horizon 2020 research and innovation program under grant agreements No. 101004719.

\section*{Data availability}

The data underlying this article will be shared on reasonable request to the corresponding author.

\bibliographystyle{mnras}
\bibliography{gaia21elv}


\appendix

\section{Photometry}
\label{sec:appendix_phot}

\begin{table*}
\centering
\caption{REM photometry of Gaia21elv.}
\label{tab:phot1}
\begin{tabular}{cccccccc}
\hline \hline
JD$\,{-}\,$2\,450\,000& $g'$& $r'$& $i'$& $z'$& $J$& $H$& $K_s$\\
\hline
9736.50&   16.91$\pm$0.10& 14.81$\pm$0.03& 13.51$\pm$0.03& 12.63$\pm$0.05& ...& ...& ...\\
9738.50&   17.02$\pm$0.24& 14.83$\pm$0.03& 13.48$\pm$0.03& 12.61$\pm$0.04& 
           9.90$\pm$0.04& 8.69$\pm$0.05& 7.79$\pm$0.04\\
9739.47&   16.84$\pm$0.08& 14.89$\pm$0.02& 13.57$\pm$0.02& 12.67$\pm$0.04& 
           9.91$\pm$0.06& 8.71$\pm$0.06& 7.79$\pm$0.09\\
9740.49&   16.93$\pm$0.11& 14.87$\pm$0.02& 13.53$\pm$0.03& 12.71$\pm$0.03& 
           9.85$\pm$0.07& 8.68$\pm$0.06& 7.82$\pm$0.02\\
9878.86&  ...& ...& ...& ...&  9.81$\pm$0.18& ...& ...\\           
9883.86&  16.87$\pm$0.09& 14.71$\pm$0.08& 13.31$\pm$0.13&  ...&  10.18$\pm$0.43&  ...& ...\\	
9889.84&  16.79$\pm$0.21& 14.66$\pm$0.05& 13.38$\pm$0.08&  ...&  ...&  ...& ...\\	
9896.75&  16.73$\pm$0.17& 14.59$\pm$0.07& 13.27$\pm$0.09&  ...&  ...&  ...& ...\\	
9901.86&  16.92$\pm$0.15& 14.67$\pm$0.09& 13.35$\pm$0.15&  ...&  ...&  ...& ...\\	
9906.86&  16.88$\pm$0.17& 14.66$\pm$0.08& 13.30$\pm$0.09&  ...&  9.70$\pm$0.49&  ...& ...\\				
9926.72&  16.79$\pm$0.21& 14.61$\pm$0.12& 13.28$\pm$0.09&  ...&  ...&  ...& ...\\
9933.72&  16.72$\pm$0.27& 14.68$\pm$0.13& 13.33$\pm$0.13&  ...&  ...&  ...& ...\\
9938.72&  16.69$\pm$0.20& 14.63$\pm$0.11& 13.31$\pm$0.10&  ...&  ...&  ...& ...\\
9943.75&  16.82$\pm$0.09&            ...& 13.37$\pm$0.02&  ...&  ...&  ...& ...\\
\hline
\end{tabular}
\end{table*}	

\begin{table*}
\centering
\caption{Photometry from other telescopes obtained for Gaia21elv.}
\label{tab:phot2}
\begin{tabular}{ccccccc}
\hline \hline
JD$\,{-}\,$2\,450\,000& $B$& $V$& $g'$& $r'$& $i'$& Telescope\\
\hline
9867.86& 18.23$\pm$0.11& 15.64$\pm$0.04& ...& 14.20$\pm$0.14& 13.04$\pm$0.08& Danish 1.54-m\\
9875.86& 18.28$\pm$0.03& 15.67$\pm$0.11& ...& 14.25$\pm$0.10& 13.07$\pm$0.10& Danish 1.54-m\\
9904.69&            ...& 15.71$\pm$0.05& ...& 14.43$\pm$0.06&            ...& PROMPT6\\
9908.70& 17.95$\pm$0.11& 15.87$\pm$0.08& ...& 14.42$\pm$0.06&            ...& UZPW 50cm\\
9909.68&            ...& 15.61$\pm$0.04& ...&            ...& 13.01$\pm$0.05& PROMPT6\\
9911.68&            ...& ...& 16.86$\pm$0.03& 14.61$\pm$0.04& 13.26$\pm$0.04& UZPW 50cm\\
9913.82&            ...& ...&            ...& 14.67$\pm$0.05& 13.35$\pm$0.06& UZPW 50cm\\
9916.84&            ...& ...&            ...& 14.67$\pm$0.05& 13.26$\pm$0.04& UZPW 50cm\\
9917.84&            ...& ...& 16.91$\pm$0.04& 14.53$\pm$0.07& 13.28$\pm$0.05& UZPW 50cm\\
9920.69&            ...& 15.65$\pm$0.06& ...& 14.39$\pm$0.07& 13.08$\pm$0.05& PROMPT6\\
9925.84&            ...& ...& 16.91$\pm$0.03& 14.69$\pm$0.04& 13.33$\pm$0.04& UZPW 50cm\\
9932.65&            ...& ...& 16.94$\pm$0.03& 14.70$\pm$0.04& 13.31$\pm$0.04& UZPW 50cm\\
9934.59&            ...& ...& 16.81$\pm$0.09& 14.66$\pm$0.04& 13.31$\pm$0.04& UZPW 50cm\\
9943.58&            ...& ...& 16.87$\pm$0.04& 14.66$\pm$0.05& 13.28$\pm$0.05& UZPW 50cm\\
9946.56&            ...& ...& 16.86$\pm$0.04& 14.71$\pm$0.03& 13.33$\pm$0.03& UZPW 50cm\\
9947.57&            ...& ...& 16.91$\pm$0.03& 14.69$\pm$0.03& 13.28$\pm$0.03& UZPW 50cm\\
9949.73&            ...& ...& 16.96$\pm$0.03& 14.72$\pm$0.05& 13.31$\pm$0.04& UZPW 50cm\\\
9950.84&            ...& ...& 16.93$\pm$0.04& 14.70$\pm$0.04& 13.31$\pm$0.03& UZPW 50cm\\
9952.71&            ...& ...& 16.91$\pm$0.04& 14.74$\pm$0.04& 13.34$\pm$0.04& UZPW 50cm\\
9954.61&            ...& 15.76$\pm$0.05& ...& 14.44$\pm$0.07& 13.16$\pm$0.06& PROMPT6\\
9955.86&            ...& ...& 16.95$\pm$0.03& 14.71$\pm$0.03& 13.34$\pm$0.05& UZPW 50cm\\
9959.66&            ...& 15.70$\pm$0.06& ...& 14.39$\pm$0.07& 13.11$\pm$0.05& PROMPT6\\
9962.64&            ...& ...& 16.91$\pm$0.03& 14.67$\pm$0.04& 13.30$\pm$0.04& UZPW 50cm\\
9963.73&            ...& ...& 16.89$\pm$0.02& 14.65$\pm$0.03& 13.28$\pm$0.04& UZPW 50cm\\
9964.63&            ...& 15.65$\pm$0.06& ...& 14.35$\pm$0.06& 13.08$\pm$0.06& PROMPT6\\
9966.87&            ...& ...& 16.88$\pm$0.03& ...& ...& UZPW 50cm\\
9969.62&            ...& 15.66$\pm$0.05& ...& 14.36$\pm$0.07& 13.09$\pm$0.05& PROMPT6\\
\hline
\end{tabular}
\end{table*}	

\end{document}